\documentclass[%
 reprint,
 amsmath,amssymb,
pra,
]{revtex4-2}

\usepackage{xcolor}
\usepackage{graphicx}
\usepackage{dcolumn}
\usepackage{bm}
\usepackage{braket}
\usepackage[normalem]{ulem}
\usepackage{amsmath}
\usepackage{textcomp}
\usepackage{tablefootnote}

\usepackage{array,booktabs}

\begin{document}


\title{Combined microwave and optical spectroscopy for  hyperfine structure analysis in thulium atoms
}

\author{D.\,Mishin}
\email{mishin.da@phystech.edu}
\affiliation{P.N.\,Lebedev Physical Institute, Leninsky prospekt 53, 119991 Moscow, Russia}

\author{D.\,Tregubov}
\affiliation{P.N.\,Lebedev Physical Institute, Leninsky prospekt 53, 119991 Moscow, Russia}



\author{N.\,Kolachevsky}
\affiliation{P.N.\,Lebedev Physical Institute, Leninsky prospekt 53, 119991 Moscow, Russia}
\affiliation{Russian Quantum Center, Bolshoy Bulvar 30,\,bld.\,1, Skolkovo IC, 121205 Moscow, Russia}

\author{A.\,Golovizin}       
\email{artem.golovizin@gmail.com}
\affiliation{P.N.\,Lebedev Physical Institute, Leninsky prospekt 53, 119991 Moscow, Russia}

\date{\today}

\begin{abstract}

The hyperfine structure of atoms and ions is widely used in fundamental and applied research. Accurate knowledge of hyperfine splitting values is essential for quantum metrology applications as well as for improving the performance of systems designed for quantum computing and simulation.
We present refined values of the hyperfine splitting frequencies of the ground $\ket{g}=\ket{4f^{13} (^2F^o)6s^2 , J = 7/2}$ and clock $\ket{c}=\ket{4f^{13} (^2F^o)6s^2 , J = 5/2}$ states of thulium atom, 
$f_{g}^\textrm{\tiny{HFS}}=1\,496\,550\,658.23(3)$\,Hz and $f_{c}^\textrm{\tiny{HFS}}=2\,113\,946\,873.08(9)$\,Hz, respectively.
The measurements are performed on an ultracold atomic ensemble in an optical lattice using combined microwave and optical transition spectroscopy.
Our results improve the accuracy by 2 and 7 orders of magnitude  for $f_{g}^\textrm{\tiny{HFS}}$ and $f_{c}^\textrm{\tiny{HFS}}$, respectively, compared to the previously published values.
We also refine the value of the Landé g-factor of the clock level to $g_c = 0.85479(11)$.

\end{abstract}

\maketitle

\section{Introduction}
\label{Section:Introduction}

Cold atom and ion ensembles have become one of the key components in many modern applications, such as quantum logic experiments\cite{dumitrescu2022dynamical,monroe2021programmable,ebadi2021quantum, graham2022multi}, quantum sensors \cite{yuan2023quantum, marciniak2022optimal}, gravimeters \cite{stray2022quantum, janvier2022compact, li2023continuous}, frequency standards \cite{boulder2021frequency,dorscher2021optical,cui2022evaluation} and others. 
All of these applications require accurate knowledge of different atomic level properties, including their energies, lifetimes, sensitivity to magnetic field, static and dynamic polarizabilities, etc.
This knowledge also facilitates the development and improvement of theoretical models \cite{glazov2019ground, ono2022observation}. 
A combination of theoretical predictions and accurate experimental data is used in fundamental studies related, for example, to precise measurements of the fundamental constants and studying their temporal variation \cite{ morel2020determination, barontini2022measuring}. 
The comparison of resonance frequencies and transition energies in different systems is used to search for dark matter \cite{flambaum2023searching, kennedy2020precision}.

Microwave transitions, typically coupling hyperfine components of the atomic level structure, are widely used in the above-mentioned applications.
They are the workhorse in frequency metrology: the definition of the SI second is now based on the microwave transition between the ground state hyperfine levels in Cs \cite{arias201850th}. 
Rb clocks \cite{ovchinnikov2011accurate} and hydrogen masers significantly contribute to the international atomic time scale TAI \cite{guena2014contributing} and form the time base in GNSS satellites \cite{hollberg2020atomic}. 
Solid-state masers operating on hyperfine transitions in various defects in diamonds are being actively studied \cite{breeze2018continuous, arroo2021perspective}.
Manipulations with microwave transitions serve in optical spectroscopy for state preparation \cite{graham2023midcircuit, sinuco2019microwave}, hyperfine averaging \cite{kaewuam2020hyperfine}, and dynamical decoupling \cite{ tan2019suppressing}.
In quantum computing and simulations, sublevels of the hyperfine structure is a common choice for a qubit encoding, and microwave transitions are used for realization of single- and two-qubit gates \cite{weidt2016trapped,valahu2021robust,srinivas2021high, wang2020high,an2022high} as well as to engineer interaction Hamiltonians\,\cite{scholl2022microwave}.

Thulium atoms have proven to be a promising platform for quantum simulators \cite{davletov2020machine, khlebnikov2019random} and transportable optical lattice clocks \cite{golovizin2021simultaneous,golovizin2021compact}.
The doublet hyperfine structure allows bicolor spectroscopy of the clock transition and the formation of the synthetic clock frequency, which is insensitive to the quadratic Zeeman shift \cite{golovizin2021simultaneous}.
Similar hyperfine splitting values of the ground $\ket{4f^{13} (^2F^o)6s^2 , J = 7/2}$ and excited $\ket{4f^{12} (^3F_4)5d_{5/2}6s^2 , J = 7/2}$ states in the optical pumping cycle open up the opportunity for the simultaneous preparation of two initial states \cite{fedorova2020simultaneous}.
Coherent excitation of the hyperfine transitions opens a new tool for state preparation and control of interatomic interactions in thulium atoms \cite{pershin2020microwave}.

Here we present the results of accurate measurements of the hyperfine splittings (HFS) of the ground $\ket{g}=\ket{4f^{13} (^2F^o)6s^2 , J = 7/2}$ and the clock $\ket{c}=\ket{4f^{13} (^2F^o)6s^2 , J = 5/2}$ states of neutral thulium atom.
In Sec.~\ref{Section:Measurement technique} we describe the measurement schemes.
The ground state HFS was measured by direct excitation of the microwave transition. 
Unlike in Ref.~\cite{pershin2020microwave}, we performed spectroscopy of the transition between sublevels with $m_F=0$, which is first-order insensitive to the magnetic field.
This provided much longer coherence time and higher measurement accuracy.  
The clock state HFS was inferred from the differential clock transition frequency and ground state HFS.
We present the HFS measurement results in Sec.~\ref{Section:Hyperfine splitting measurements}, and analyse the systematic shifts and uncertainties in Sec.~\ref{Section:Systematic shift analysis}.
In Sec~\ref{Section:Measurement of the clock level g-factor} we discuss the clock level Landé g-factor measurement.
The conclusions are drawn in Sec.~\ref{Section:conclusion}.

\section{Measurement techniques}
\label{Section:Measurement technique}

\begin{figure}[ht!]
\center{
\resizebox{0.45\textwidth}{!}{
\includegraphics{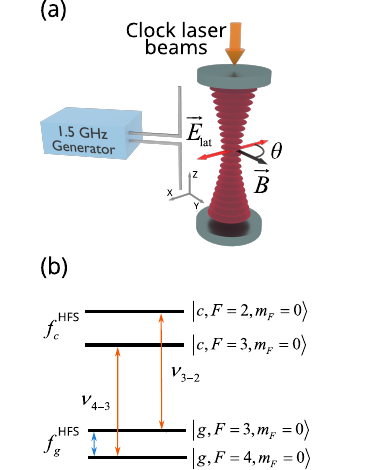}
}
\caption{Experimental setup (a) and simplified level structure of the ground and clock states (b).
Hereinafter blue colors on the figures correspond to the microwave transition and orange colors correspond to the optical clock transitions.
}
\label{fig:Lattice and levels}}
\end{figure}

The procedure for preparing an ensemble of cold thulium atoms in a one-dimensional optical lattice is similar to our previous works \cite{golovizin2021simultaneous, 506_surer_cooling}.
After the two-stage laser cooling atoms are trapped into the near-magic wavelength optical lattice at 1063.6\,nm with a typical depth of $\sim1200$\,$E_\textrm{rec}$.
Then we perform optical pumping to $\ket{g,F=4, m_F = 0}$ and $\ket{g,F=3, m_F = 0}$ states on the 418\,nm transition \cite{mishin2022effect,fedorova2019optical, fedorova2020simultaneous}.
If necessary, we can depopulate the $\ket{g,F=3, m_F = 0}$ state using repumping radiation during the optical pumping step.
Since optical pumping results in the heating of the trapped atoms, we subsequently reduce the lattice depth down to $\sim100$\,$E_\textrm{rec}$ for 20\,ms, allowing the hottest atoms to leave the spectroscopy region \cite{mishin2022effect}.
After this step, we can raise the depth of the optical lattice to the desired level, which is $200$\,$E_\textrm{rec}$ unless specified otherwise.
The readout technique (described in detail in \cite{golovizin2021simultaneous}) allows us to measure atomic population in all 4 states used in the current experiment [see Fig.~\ref{fig:Lattice and levels}(b)] in one experimental cycle.
Unless specified otherwise, we set the total number of atoms at 5000 for microwave transition spectroscopy experiments and 8000 for optical transition spectroscopy experiments.

To excite the hyperfine transition between the ground state sublevels we use a dipole antenna adjusted to a $\sim$$1.5$\,GHz resonant frequency and driven by the SRS\,SG382 signal generator.
Depending on the vacuum chamber design, this antenna can be located either near a viewport or inside the vacuum chamber in close proximity to the atomic cloud.
For microwave  spectroscopy  we use the Ramsey technique: two Ramsey $\pi/2$ pulses are applied with a duration of $\tau\sim1$\,ms separated by a free-evolution time of $T=0.5$\,s. 
We record Ramsey spectra by scanning the microwave (MW) frequency in the $\pm2$\,Hz region around the resonance.
To isolate the central Ramsey fringe, we gradually increase $T$ from 1\,ms to 500\,ms.
We perform such measurement in two configurations, when the bias magnetic field is perpendicular ($\theta=\pi/2$) and parallel ($\theta=0$) to the optical lattice polarization [see Fig.~\ref{fig:Lattice and levels}(a)].

The measurement of the clock level hyperfine splitting is done using a previously developed bicolor interrogation technique of two clock transitions \cite{golovizin2021simultaneous}.
Our experimental setup allows us to independently stabilize frequencies of two laser fields that excite them.
The frequencies of these fields are denoted by $\nu_{4-3}$ and $\nu_{3-2}$ respectively  [see Fig.~\ref{fig:Lattice and levels}(b)].
The difference between them is designated as the differential frequency 
\begin{eqnarray} \label{eq:nu_d}
    \nu_\textrm{d} \equiv \nu_{\scriptscriptstyle 3-2} - \nu_{\scriptscriptstyle 4-3} = f_{c,{\scriptscriptstyle 0-0}} -f_{g,{\scriptscriptstyle 0-0}},
\end{eqnarray} 
where $f_{g(c),{\scriptscriptstyle 0-0}}$ is the frequency of the ground(clock) state hyperfine transition.
Thus, combined microwave and optical spectroscopy allows us to measure the hyperfine splitting frequencies of both the ground and clock states.
We note that differential clock frequency measurements were performed only at $\theta=\pi/2$.
This is because the magic wavelength of the optical lattice, which is required for the narrow-line optical spectroscopy, is realized at 1063.6\,nm and $\theta=\pi/2$ \cite{golovizin2019inner, mishin2022effect}.
When $\theta=0$, the clock transition frequency shift and broadening due to the dynamic Stark effect in our experiment are at the kHz level, which makes precision optical spectroscopy impossible.


\section{Hyperfine splitting measurement}
\label{Section:Hyperfine splitting measurements}

\begin{figure*}[ht!]
\center{
\resizebox{\textwidth}{!}
{
\includegraphics{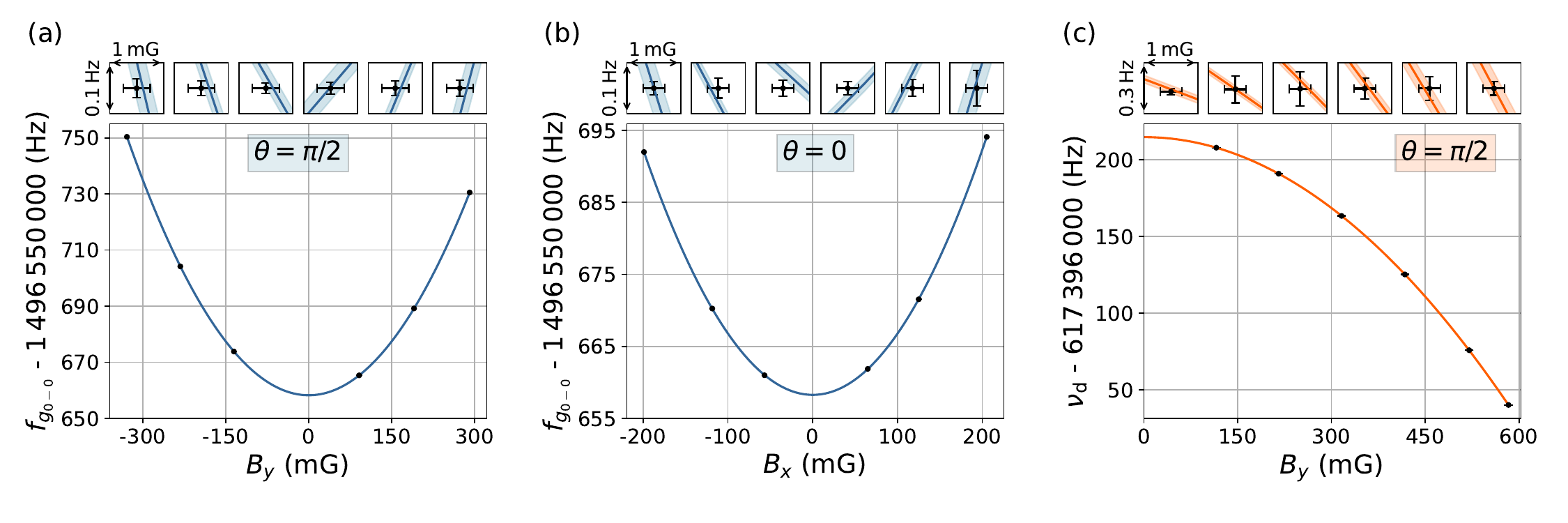}
}\\

\caption{ Second order Zeeman effect measurements. 
Hyperfine splittig measurements for $\theta = \pi/2$ (a) and $\theta = 0$ (b) and differential clock transitions frequency measurements for $\theta = \pi/2$ (c) are presented.
Solid lines are the parabolic fits.
Insets show zoom-ins to the data points (black crosses) and fit error (colored strip) with vertical and horizontal scales indicated near the first inset.}
\label{fig:Parabolas}}
\end{figure*}

The bias magnetic field in the experiment is necessary for resolving  the magnetic sublevels.
We measured transition frequencies at different magnetic fields and extrapolated results to the zero value.
The spin of the thulium nucleus is $I=1/2$, so the hyperfine structure energy levels in the presence of a magnetic field can be described by the Breit-Rabi formula \cite{giglberger1967ground, sukachev2016inner}.
Deriving  the shift up to the second order of the magnetic field, we obtain
\begin{eqnarray}
    \Delta &&E_{F=J\pm\frac{1}{2}, m_F}(B) = m_F\left(g_J\mu_\textrm{\tiny{B}} \mp \frac{(g_J\mu_\textrm{\tiny{B}}-g_I\mu_\textrm{\tiny{N}})}{(2J+1)}\right)B\nonumber\\
    &&\pm \frac{(g_J\mu_\textrm{\tiny{B}}-g_I\mu_\textrm{\tiny{N}})^2}{h A_J(2J+1)} \left( \frac{1}{2} - \frac{2m_F^2}{(2J+1)^2}\right)B^2\,,
    \label{eq:BreitRabiX2}
\end{eqnarray}
where $g_J$ and $g_N$ are the electron and nuclear Landé g-factors and $\mu_\textrm{\tiny{B}}$ and $\mu_\textrm{\tiny{N}}$ are the Bohr and the nuclear magnetons, $A_J$ is the hyperfine splitting constant of a particular level. 
The first-order Zeeman shift is zero for the $m_F=0$ sublevel.
The microwave transition frequency between the central magnetic sublevels of the ground state is then
\begin{eqnarray}
    f_{g,{\scriptscriptstyle 0-0}} = f_{g}^\textrm{\tiny{HFS}} + \frac{(g_J\mu_\textrm{\tiny{B}}-g_I\mu_\textrm{\tiny{N}})^2}{h^2 A_J(2J+1)} B^2 = f_{g}^\textrm{\tiny{HFS}} + \beta_g B^2\,,
     \label{eq:quadZeeman}
\end{eqnarray}
where $f_{g}^\textrm{\tiny{HFS}}$ is the ground state hyperfine splitting frequency  including corrections introduced by other effects discussed below, and $\beta_{{g}} \approx 852\,\textrm{Hz/G}^2$ is the quadratic Zeeman coefficient of the $\ket{g, F=4, m_F=0} \rightarrow \ket{g, F=3, m_F=0}$ transition.
Magnetic field was calibrated before and after each measurement. 
The calibration was done via spectroscopy of $\sigma^{\pm}$ clock transitions $\ket{g, F=4, m_F=0}\rightarrow\ket{c, F=3, m_F=\pm1}$, which are first-order sensitive to the magnetic field.
We determine the magnetic field from the frequency difference between $\sigma^+$ and $\sigma^-$ transitions, for which the quadratic Zeeman and tensor Stark shifts are canceled.
Calibration results also show that the magnetic field fluctuations are less than 0.2\,mG during the time of the experiment.

Experimental data on $f_{g,{\scriptscriptstyle 0-0}}(B)$ and parabolic fits by Eq.~(\ref{eq:quadZeeman}) are shown in Fig.~\ref{fig:Parabolas}(a,\,b) for two orientations of the bias magnetic field. 
Zoom-ins near the experimental points with 100\,mHz vertical and 1\,mG horizontal scales are shown above the main plot. 
Vertical error bars correspond to the frequency measurement uncertainty (statistical error of the Ramsey fringe line center determination), and horizontal error bars represent the estimated magnetic field fluctuation over the course of the measurement. 
We find $f_{g}^\textrm{\tiny{HFS}}(\theta=\pi/2, B=0) = 1\,496\,550\,658.259(32)$\,Hz and $f_{g}^\textrm{\tiny{HFS}}(\theta=0, B=0) = 1\,496\,550\,658.269(34)$\,Hz, where the uncertainty includes both the fit error and the magnetic field fluctuation. 



Eq.~(\ref{eq:BreitRabiX2}) is also  valid for the upper clock level.
The deferential frequency $\nu_\textrm{d}$ [see Eq.~(\ref{eq:nu_d})] depends on the magnetic field as
\begin{eqnarray}\label{eq:quad_Zeeman_diff}
    \nu_\textrm{d}  = f_{c,{\scriptscriptstyle 0-0}} -f_{g,{\scriptscriptstyle 0-0}} =f_{c}^\textrm{\tiny{HFS}} - f_{g}^\textrm{\tiny{HFS}} + (\beta_c - \beta_g) B^2    
\end{eqnarray}
with $\beta_c \approx 338\,\textrm{Hz/G}^2$.
Similarly to the microwave frequency measurement, the magnetic field was calibrated before and after every measurement.

Measurement data $\nu_\textrm{d} (B)$ and the parabolic fit are shown in Fig.~\ref{fig:Parabolas}(c).
For each measured point the laser frequency stabilization to the clock transitions in thulium atoms was performed. 
The measurement error is largely determined by the instability of the external magnetic field over the measurement.
As a compromise between reducing the measurement uncertainty (typically 50\,mHz for the differential frequency) and long-term magnetic field drifts (typically $0.2$\,mG), the  measurement time interval was chosen to be 15\,minutes.
From the fit we obtain $\nu_\textrm{d}(B\!=\!0) = 617\,396\,214.818(31)$\,Hz.

\section{Systematic shift analysis}
\label{Section:Systematic shift analysis}
Here, we present a detailed study of the shifts and associated uncertainties caused by systematic effects in our experiment.
We analyze these effects separately from the measurements performed earlier to evaluate the corresponding shifts and apply corrections, thereby achieving unperturbed hyperfine splitting values.

\subsection{Collisional shift}
\label{subsection:Collisional shift}

\begin{figure*}[ht!]
\center{
\resizebox{0.7\textwidth}{!}{
\includegraphics{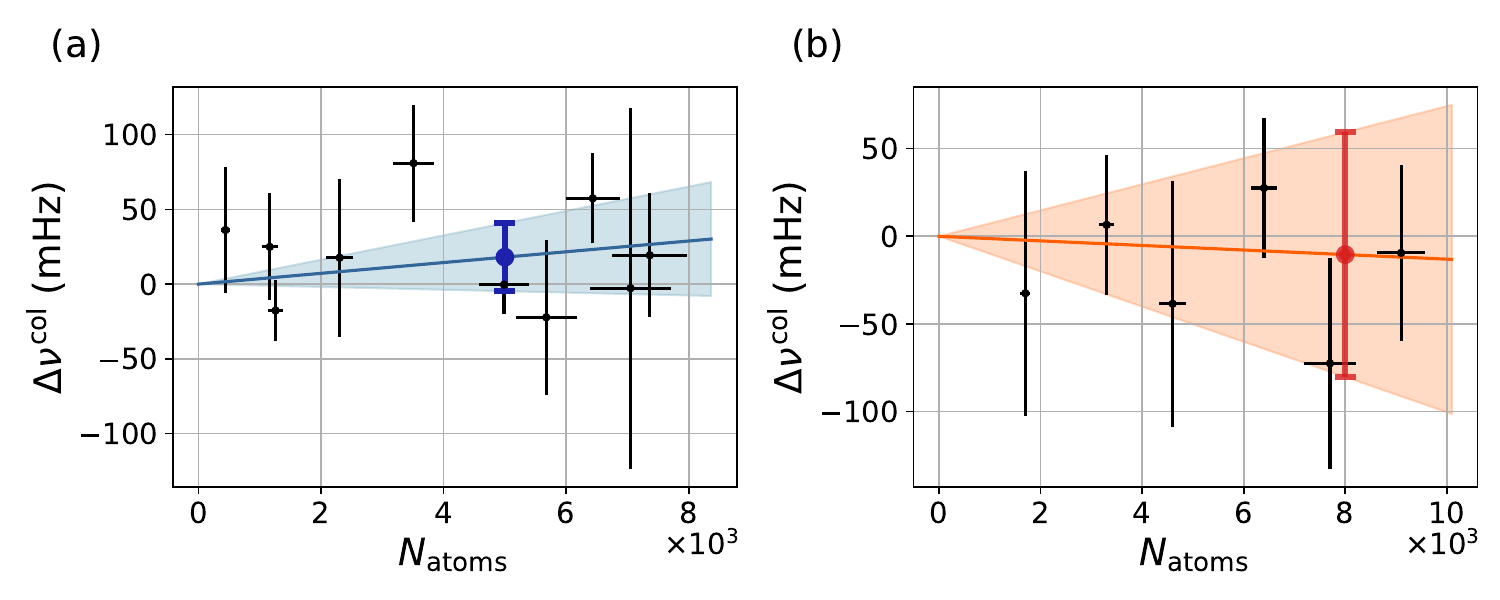}
}\\

\caption{Collisional shift for the microwave transition frequency (a) and the differential frequency of two clock transitions (b).
Vertical error bars correspond to the achieved frequency determination error while horizontal error bars correspond to one standard deviation of the total number of atoms detected in each experiment.
The data is fitted with linear function and shifted by the offset frequency value $\nu_0$, see Eq.~(\ref{eq:dens}) and the discussion in the text.
The colored strip illustrates 1\,s.d. fit uncertainty of the slope value.
The blue and red points illustrate the collisional frequency shift and uncertainty at the operational values used in Sec.~\ref{Section:Hyperfine splitting measurements}.}
\label{fig:Collisions}}
\end{figure*}

To estimate the collisional shift of $f_{g}^\textrm{\tiny{HFS}}$ and 
$\nu_\textrm{d}$ from Sec.~\ref{Section:Hyperfine splitting measurements}, we have measured the dependence of both microwave and differential clock transitions frequencies as a function of the number of atoms ranging between $10^3$ and $10^4$ at $B_0 = 200$\,mG along the Y axis ($\theta=\pi/2$).
The number of atoms was changed during the state preparation step, so it did not impact other parameters of the experiment, including the size of the atomic cloud.
Assuming a linear dependence of the frequency on the number of atoms $N$ due to the low-density regime ($\sim10$ atoms per optical lattice cell), the measured frequency value and the corresponding collisional shift can be expressed as
\begin{subequations}
\begin{equation}
    \nu(B_0, N) = \nu_0(B_0) + kN, \label{eq:dens_a}
\end{equation}
\begin{equation}
    \Delta\nu^{\textrm{\tiny{col}}}(N)  = \nu(B_0,N) - \nu_0(B_0) = kN,\label{eq:dens_b}
\end{equation}
    \label{eq:dens}
\end{subequations}
where $k$ is the slope and $\nu_0(B_0)$ is the frequency at $N\rightarrow0$, i.e. unperturbed by the effect under consideration.
Experimental data (black points with error bars) and approximations with Eq.~(\ref{eq:dens_a}) (solid line) are shown on Fig.~\ref{fig:Collisions}.
For visualization clarity, we have subtracted $\nu_0$, obtained from the fit, from the measured frequencies, hereby the solid line follows Eq.~(\ref{eq:dens_b}). 
The shaded area depicts 1\,s.d. fit uncertainty of the slope value $k$.
The blue and red points with error bars denote the collisional shift and uncertainty of the measured transition frequencies at the operational number of atoms in measurements described above in  Sec.~\ref{Section:Hyperfine splitting measurements}.
The resulting shift (uncertainty) is $\Delta \nu_{g}^{\textrm{\tiny{col}}} = 18 (23)$\,mHz for the ground level hyperfine splitting and $\Delta \nu_\textrm{d}^{\textrm{\tiny{col}}} = -9 (71)$\,mHz for the differential clock frequency. 
One can see that the shift is compatible with zero within the measurement uncertainty.  

\subsection{State mixing shift}
\label{subsection:State mixing shift}

The simultaneous interaction of an atom with magnetic and ac-electric fields  can lead to a change of the eigenstates of the Hamiltonian, in particular, to the mixing of the $m_F = 0$  magnetic sublevel  with $m_F \neq 0$ ones. 
This can result in a frequency shift denoted below as the state mixing shift.
The contribution of this effect depends on the angle $\theta$ between the bias magnetic field and the polarization of the optical lattice [see Fig.~\ref{fig:Lattice and levels}(a)], as well as on magnitudes of the bias magnetic field and lattice depth. 
The state mixing shift at $\theta = 0$ becomes zero because quantisation axes coincide and both effects lead to the same eigenstates.  
In turn, for $\theta = \pi/2$ this effect can lead to a significant frequency shift\,\cite{virgo_simultaneous_2013,STATE_mixing}.

Since the state mixing shift depends on both the lattice depth and magnetic field, we calculated the corresponding shift value for each experimental point shown in Fig.~\ref{fig:Parabolas}.
It is worth noting that owing to the low optical lattice depth $u=200\,E_\textrm{rec}$, maximum corrections were 6\,mHz for the ground state hyperfine spectroscopy and 21\,mHz for the differential clock transitions frequency  measurements.
We fitted the corrected data using the same parabolic dependencies [Eqs.~(\ref{eq:quadZeeman}-\ref{eq:quad_Zeeman_diff})] and assigned the state mixing shift a value equal to the difference between the frequency values obtained in Sec.~\ref{Section:Hyperfine splitting measurements} and the corresponding values derived from the corrected data fits. 
For the $\theta = \pi/2$ experiments state mixing shifts are $\Delta \nu_{g}^{\textrm{\tiny{s.m.}}} = 5$\,mHz for the ground level HFS frequency and $\Delta \nu_\textrm{d}^{\textrm{\tiny{s.m.}}} = 20$\,mHz for the clock transition differential frequency.
As a conservative estimation, we assign the uncertainty of these shifts to be equal to the shift magnitudes.
For the ground state splitting measurements at $\theta = 0$ we consider state mixing shift to be zero, and assign $\delta \nu_{g}^{\textrm{\tiny{s.m.}}} = 1$\,mHz uncertainty due to possible misalignment $\delta\theta \lesssim 0.1$\,rad of the magnetic field and lattice polarization.

\subsection{Stark shift}
\label{subsection:Stark shift}

\begin{figure*}[ht!]
\center{
\resizebox{\textwidth}{!}{
\includegraphics{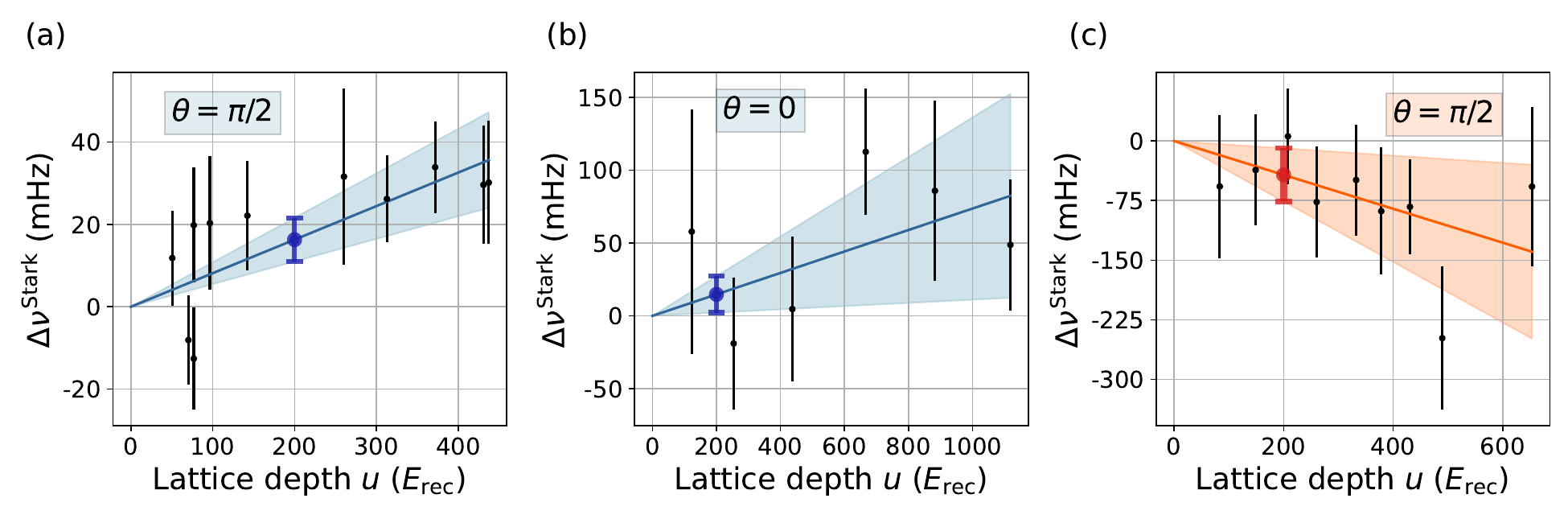}
}\\

\caption{
The dependence of the Stark shift on the optical lattice depth.
Data for the microwave transition frequency for $\theta = \pi/2$ and $\theta = 0$ experiments are presented in  (a) and  (b), respectively, while experimental data for the differential frequency of two clock transitions for $\theta = \pi/2$ is presented in  (c).
Error bars represent the frequency measurement uncertainties.
Similar to the collisional shift measurements, experimental data are fitted with a linear function (solid lines) and shifted by an offset frequency value.
The colored stripe illustrates 1 s.d. fit uncertainty of the slope.
The blue and red points illustrate the Stark frequency shift and uncertainty at the operational lattice depth values used in Sec.~\ref{Section:Hyperfine splitting measurements}.}
\label{fig:Stark}}
\end{figure*}

Polarizabilities of the $m_F = 0$ sublevels in one hyperfine doublet are  equal in $JI$ coupling approximation\,\cite{sukachev2016inner}.
However, the next-order corrections of the electron-nucleus interaction may result in a non-zero difference in polarizability between these states. 
This would lead to a linear dependence of the transition frequency on the optical lattice depth due to the dynamic Stark effect. 
To characterize this effect, we measured transition frequencies as a function of the optical lattice depth while the magnetic field was set to  $B=200$\,mG for the microwave transition experiments, and $B=300$\,mG for the differential clock transitions frequency experiments. 
Since both the state mixing and Stark shifts depend on the angle $\theta$, the measurements were done for configurations at $\theta = \pi/2$ and $\theta = 0$ for the ground state hyperfine transition.
Corrections caused by the state mixing were introduced for each experimental point.
In order to minimize the influence of the state mixing, the depth of the optical lattice in experiments at $\theta = \pi/2$ was limited to $600\,E_\textrm{rec}$.
In this case, the corrections from the state mixing do not exceed the statistical error of the experiment.
The data analysis here is similar to the collisional shift estimation described above: the data is fitted with a linear function with two free parameters.
Results of the Stark shift measurements are shown in Fig.~\ref{fig:Stark}:  the solid lines depict the linear fit results, and the colored strips correspond to $1\,\sigma$ fit errors of the slope. 
Experimental points and linear fits are shifted by the obtained offset frequency value for clarity of visualization.
The blue and red points illustrate the Stark frequency shift and uncertainty at operational lattice depth values used in Zeeman shift measurements in Sec.~\ref{Section:Hyperfine splitting measurements}.
The resulting Stark shifts are $\Delta \nu_{g}^{\textrm{\tiny{Stark}}} = 16 (5)$\,mHz for $\theta = \pi/2$ 
and $\Delta \nu_{g}^{\textrm{\tiny{Stark}}} = 14 (12)$\,mHz for $\theta = 0$ ground state splitting measurements, and $\Delta \nu_\textrm{d}^{\textrm{\tiny{Stark}}} = -43 (33)$\,mHz for $\theta = \pi/2$ differential clock frequency experiment.
The measured slopes correspond to the polarizability difference of  $\Delta \alpha \sim 10^{-5}$ in atomic units.
We note that no noticeable changes in the atomic cloud size were observed for different lattice depths, so we consider the density shift to be constant during these experiments.

\subsection{Other systematic shifts}
\label{subsection:Other systematic shifts}

Typical values of the blackbody radiation shift for hyperfine transitions are at the level of $10^{-15}$\,rel.u., which is much lower compared to the accuracy of the present experiments. 
To estimate this shift for thulium atoms, we use the measured value of the differential polarizability at the optical lattice wavelength of 1063.6\,nm. 
For $\Delta\alpha\sim10^{-5}$ in atomic units the BBR shift at $300$\,K is $\delta\nu_\textrm{BBR}\sim10^{-7}$\,Hz, or below $10^{-16}$ in relative units.

The first-order Doppler shift in the current configuration is suppressed due to the Lamb-Dicke regime, both for microwave and clock transition spectroscopy.
The second order Doppler shift is at $10^{-20}$ relative level for 10\,\textmu K temperature of atomic cloud.

The effects detailed in this subsection have a minor impact relative to those described previously; therefore, they will be omitted from the analysis of final values and uncertainties.

\subsection{Final values of the hyperfine splittings}

The error budget and experimental results can be seen in Table~\ref{tab:my_label}.
For each effect considered we present corresponding correction that should be applied to the value measured in Sec.~\ref{Section:Hyperfine splitting measurements} to account for the effect.
The final value of the ground state hyperfine splitting is calculated from the weighted average of the two measurements at $\theta=\pi/2$ and $\theta=0$, both corrected for the systematic frequency shifts:
\begin{eqnarray} \label{eq:gnd_hfs_final}
    f_{g}^\textrm{\tiny{HFS}}  = 1\,496\,550\,658.228(33)\,\textrm{Hz}, 
\end{eqnarray}
where we take into account that the density shift is common for the two configurations.
The clock state HFS is calculated from the measured differential clock frequency and the ground state HFS  using Eq.~(\ref{eq:nu_d}) and value from Eq.~(\ref{eq:gnd_hfs_final}).
When calculating the final uncertainty, we assumed the errors from the considered effects, as well as the measurements of $\nu_\textrm{d}$ and $f_{g}^\textrm{\tiny{HFS}}$, to be independent.
\begin{eqnarray}
    f_{c}^\textrm{\tiny{HFS}} =
    2\,113\,946\,873.078(93)\,\textrm{Hz}.
    \label{eq:corrections_clock}
\end{eqnarray}

\begin{table*}
\caption{\label{tab:my_label}    
Hyperfine splittings measurement results.
The measured uncertainty is calculated from the fit uncertainty of the transition frequency at $B=0$ and includes both frequency and magnetic field measurement errors  (see Fig.~\ref{fig:Parabolas}).
The shifts and associated uncertainties are stated for $u=200\,E_r$ and $N_\textrm{atoms}=5\times10^3$ ($N_\textrm{atoms}=8\times10^3$) measurement configuration for the ground state microwave transition frequency (differential clock frequency).
The final value of the ground state HFS is obtained from the weighted average of the two measurements at $\theta=\pi/2$ and $\theta=0$, with the common uncertainty from the density shift taken into account.
The value of the clock state HFS is calculated from Eq.~(\ref{eq:nu_d})}
\begin{ruledtabular}
\begin{tabular}{c@{\hskip 0.4in}cccc@{\hskip 0.4in}cc}
 &\multicolumn{2}{c}{$f_{g}^\textrm{\tiny{HFS}}$, $\theta = \pi/2$}&\multicolumn{2}{c}{$f_{g}^\textrm{\tiny{HFS}}$, $\theta = 0$}&\multicolumn{2}{c}{$\nu_\textrm{d} = f_{c}^\textrm{\tiny{HFS}} - f_{g}^\textrm{\tiny{HFS}}$, $\theta = \pi/2$}\\
 \hline
 & Value, Hz & $1\,\sigma$, mHz & Value, Hz & $1\,\sigma$, mHz & Value, Hz & $1\,\sigma$, mHz \\
 \hline
 Measured  &  $1\,496\,550\,658.259$  & 32 & $1\,496\,550\,658.269$ & 34 & $617\,396\,214.818$ & 31 \\
 \hline
           Stark shift correction&   -0.016  & 5 & -0.014 & 12 & 0.043 & 33 \\
           Density shift correction&   -0.018  & 23 & -0.018 & 23 & 0.009 & 71 \\
           State mixing correction&   -0.005  & 5 & 0 & 1 & -0.020 & 20 \\
\hline
Corrected &  $1\,496\,550\,658.221$  & 40 & $1\,496\,550\,658.237$ & 43 & $617\,396\,214.850$ & 87 \\
\hline
{\bf Final} & \multicolumn{4}{c}{$f_{g}^\textrm{\tiny{HFS}}=1\,496\,550\,658.228(33)$\,Hz} & \multicolumn{2}{c}{$f_{c}^\textrm{\tiny{HFS}}=2\,113\,946\,873.078(93)$\,Hz} \\
\end{tabular}
\end{ruledtabular}
\end{table*}

\section{Measurement of the clock level g-factor}
\label{Section:Measurement of the clock level g-factor}

As mentioned above, the convenient calibration of the bias magnetic field during the optical clock operation can be done by spectroscopy of $\sigma^\pm$ transitions from the $\ket{g, F=4, m_F=0}$ sublevel.
Thus, higher accuracy of the Landé g-factor $g_c$ of the clock state would improve the calibration precision.
As the the ground state g-factor $g_g=1.141189(3)$ is known with high accuracy \cite{giglberger1967ground}, interrogation of the clock transitions between different magnetic sublevels allows us to refine the value of clock state g-factor $g_c=0.855(1)$ published previously in \cite{blaise1965progres}.
For this measurement, we skip the optical pumping step, so that atoms are distributed over all magnetic sublevels of the ground state.

\begin{figure}[ht!]
\center{
\resizebox{0.45\textwidth}{!}{
\includegraphics{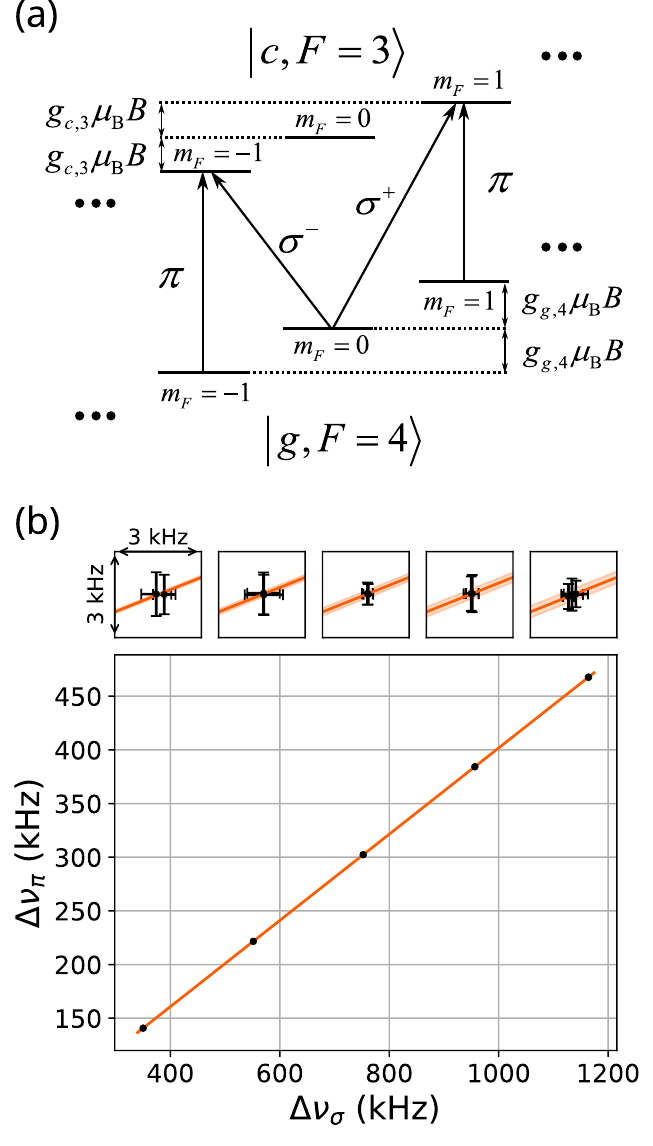}
}\\

\caption{ Measurements of the clock level Landé g-factor. a) Scheme of the used transitions with corresponding linear Zeeman shifts notation.
b) Measured $\Delta \nu_{\pi}$ (vertical axis) and $\Delta \nu_{\sigma}$ (horizontal axis), see Eqs.~(\ref{eq:sigma}) in the main text, ratio is fitted with the linear function.
The colored stripe illustrates 1\,s.d. fit uncertainty of the slope.
Insets above the main graph show zoom-ins to the experimental points with 3 kHz scales both for the horizontal and vertical axes.}
\label{fig:g-factor}}
\end{figure}

In order to determine $g_c$, we performed frequency measurements of 4 clock transitions [see Fig.~\ref{fig:g-factor}(a)]: $\nu_{0, \sigma^+}$ and $\nu_{0, \sigma^-}$ of $\sigma^\pm$ transitions from the $\ket{g, F=4, m_F=0}$ sublevel, and $\nu_{+1,\pi}$ and $\nu_{-1,\pi}$ $\pi$ transitions from the $\ket{g, F=4, m_F=+1}$ and $\ket{g, F=4, m_F=-1}$ sublevels, respectively.
We scanned the clock laser frequency through the corresponding resonances.
To infer the excitation efficiency, we measured the number of atoms in the $\ket{g, F=4}$ state twice. 
The first measurement was done right after the clock pulse, giving us the number of non-excited atoms. The second measurement was done 300\,ms later, which is longer than the clock level natural lifetime $\tau=112$\,ms, so the majority of the excited atoms had spontaneously decayed.

Differences between the corresponding frequencies  
\begin{subequations}\label{eq:sigma}
\begin{equation}
    \Delta \nu_{\sigma} = \nu_{0,\sigma^+}-\nu_{0,\sigma^-}= 2g_{c,3} \mu_\textrm{\tiny{B}} B \label{eq:sigmapm}
\end{equation}
\begin{equation}
    \Delta \nu_{\pi} = \nu_{-1,\pi}-\nu_{+1,\pi}= 2(g_{g,4}-g_{c,3}) \mu_\textrm{\tiny{B}} B \label{eq:sigmapi}
\end{equation}
\end{subequations}
are determined by the first-order Zeeman shifts because both the tensor Stark shift from the optical lattice \cite{sukachev2016inner} and the quadratic Zeeman shift [Eq.~(\ref{eq:BreitRabiX2})] are canceled in each difference (due to their proportionality to $m_F^2$).
Here $g_{g,4} = g_g\mu_\textrm{\tiny{B}} - (g_g\mu_\textrm{\tiny{B}}-g_I\mu_\textrm{\tiny{N}})/8$ and $g_{c,3} = g_c\mu_\textrm{\tiny{B}} - (g_c\mu_\textrm{\tiny{B}}-g_I\mu_\textrm{\tiny{N}})/6$ are derived from Eq.~(\ref{eq:BreitRabiX2}).
From the ratio of Eq.~(\ref{eq:sigmapm}) and Eq.~(\ref{eq:sigmapi}) one can find 
\begin{eqnarray} \label{eq:gc3}
    g_{c,3}=g_{g,4} \frac{1}{1+\Delta\nu_\pi/\Delta\nu_\sigma}\,,
\end{eqnarray}
which eliminates the absolute value of the magnetic field from the equation. 

The measurements were performed at different bias magnetic fields $B_0=200\div600$\,mG.
Experimental data of $\Delta \nu_{\pi}(\Delta \nu_{\sigma})$ is shown with dots in Fig.~\ref{fig:g-factor}(b), while the solid line depicts the linear fit with the slope as a single fit parameter. 
Using Eq.~(\ref{eq:gc3}) we find from the fit $g_{c,3} = 0.71233(10)$.
This allows us to  infer $g_c = 0.85479(11)$, which is in agreement with the previously published value \cite{blaise1965progres} but is 10 times more accurate.
The main uncertainty comes from the transition line center determination error, which is mostly associated with magnetic field fluctuations and spatial inhomogeneity.
Contribution from uncertainties of both $g_g$ and $g_I$ values that were used in calculations are at least two orders smaller than experimental error. 

\section{Conclusion}
\label{Section:conclusion}
Using combined microwave and optical spectroscopy techniques, we have refined the values of the ground and clock states hyperfine splittings $f_{g}^\textrm{\tiny{HFS}}=1\,496\,550\,658.23(3)$\,Hz (relative error $3\times10^{-11}$) and $f_{c}^\textrm{\tiny{HFS}}=2\,113\,946\,873.08(9)$\,Hz ($6\times10^{-11}$) in thulium atom, which is an improvement of $250$ and $10^7$ times compared to the previously published results  \cite{giglberger1967ground, van1980high}. 
The uncertainty is primarily limited by the stability of the bias magnetic field of $\sim0.2$\,mG.
Selective excitation of the clock transitions between different magnetic sublevels allowed us to refine the clock level Landé g-factor  $g_{c} = 0.85479(11)$ with the main uncertainty also associated with the magnetic field fluctuations.
Both these results can be further improved by the use of magnetic shields or active stabilization of the bias magnetic field.

Our results open the possibility for further improvement of the performance of thulium optical clocks.
The obtained estimate of the collisional shift is compatible with zero for both the hyperfine transition and the clock transitions' differential frequency.
In order to characterize this effect better, it is necessary either to significantly increase the number of atoms during interrogation or to increase the concentration of atoms by reducing the cloud size.
The latter can be achieved using deep laser cooling \cite{506_surer_cooling}, as the characteristic size of the cloud in an optical lattice is determined by the temperature of the atoms.
Estimated differential polarizabilities of the ground and clock state hyperfine transitions are of the same order of $10^{-5}$ in atomic units, which means that the polarizabilities of two clock transitions are close to each other.
This result supports the applicability of the bicolor interrogation scheme \cite{golovizin2021simultaneous} for realization of a high-accuracy optical clock, which requires the lattice wavelength to be magic for both clock transitions simultaneously.
Finally, during the measurements reported here, excitation of Ramsey oscillations with a free evolution time of 0.5\,seconds was used, while a coherence time on the order of several seconds was observed \cite{COHERENCE_TIMES}.
Such coherence time makes the ground state hyperfine transition in thulium an interesting candidate for quantum logic and quantum memory applications.

\begin{acknowledgments}

The authors acknowledge the support of RSF grant no. 23-22-00437 . 

\end{acknowledgments}

\bibliography{biblio}

\end{document}